\documentclass[prd,twocolumn,showpacs,floatfix,amsmath,nofootinbib,amssymb,floatfix]{revtex4}
\usepackage{graphicx,color,dcolumn,booktabs,bm}
\usepackage{longtable,lscape}
\usepackage{txfonts}
\usepackage{overpic}
\usepackage{amssymb}
\usepackage{indentfirst}
\usepackage{feynmf}   %{feynmp}
\usepackage{slashed}  %for Feynman symbols
\usepackage{cases}
\usepackage{color}
\usepackage{multirow}
\usepackage{epstopdf}
\usepackage{longtable}
\usepackage{graphicx,color,dcolumn,booktabs,bm}
\usepackage[colorlinks,
            citecolor=blue,
            anchorcolor=red,
            menucolor=red,
            linkcolor=red,
            filecolor=red,
            runcolor=red,
            urlcolor=blue,
            frenchlinks=red]{hyperref}

\graphicspath{{Figures/}} %

\allowdisplaybreaks

\begin{document}
%\begin{CJK}{GBK}{}

\title{Heavy molecules and one-$\sigma/\omega$-exchange model}

\author{Rui Chen$^{1,2,3}$}
\email{chenr15@lzu.edu.cn}
\author{Atsushi Hosaka$^{3}$}
\email{hosaka@rcnp.osaka-u.ac.jp}
\author{Xiang Liu$^{1,2}$}
\email{xiangliu@lzu.edu.cn}

\affiliation{ $^1$School of Physical
Science and Technology, Lanzhou University, Lanzhou 730000, China
\\
$^2$Research Center for Hadron and CSR Physics, Lanzhou University
and Institute of Modern Physics of CAS, Lanzhou 730000, China\\
$^3$Research Center for Nuclear Physics (RCNP), Osaka University, Ibaraki, Osaka 567-0047, Japan}

\begin{abstract}
In the framework of the one-boson-exchange model, we explore whether the intermediate- and short-range forces from $\sigma/\omega$ exchange can be strong enough to bind heavy molecular states. $\Lambda_cD(\bar{D})$, and $\Lambda_c\Lambda_c(\bar{\Lambda}_c)$ systems have been studied and compared. We find that the force from $\sigma$ exchange is attractive and dominant, whereas the $\omega$-exchange force is not. As a consequence, the S-wave $\Lambda_cD$, $\Lambda_c\Lambda_c$, and $\Lambda_c\bar{\Lambda}_c$ can be possible molecular candidates. We further indicate that a one hadron-hadron system with more light quarks $(u,d)$ can be easier to form a bound state. As a by-product, by studying the heavy-quark mass dependence for the $\Lambda_cD(\bar{D})$-like and $\Lambda_c\Lambda_c(\bar{\Lambda}_c)$-like systems, we find that the charm/bottom sector can easily accommodate molecular states. Finally, the $\Lambda_cN(\bar{N})$ and $\Lambda_bN(\bar{N})$ systems are investigated. Our results indicate that they are also likely to form bound states. By including one-$\pi$-exchange forces providing additional attraction when coupled channels are included, we expect many molecular states in heavy-quark sectors.
\end{abstract}

%\keywords{Molecular State,
%Exotic State, One Pion Exchange, Effective Potential}

\maketitle

\section{introduction}\label{sec1}

{{Ever since the observations of $X(3872)$ \cite{Choi:2003ue} and $\Theta^+$ \cite{Nakano:2003qx}, much evidence has been reported for new types of structures that are beyond the minimal $\bar q q$ mesons and $qqq$ baryons, although the observation of $\Theta^+$ have been criticized by subsequent experiments~\cite{Battaglieri:2005er,McKinnon:2006zv}.}}
Hence, they are called the exotic hadrons, like $X(3872)$ \cite{Choi:2003ue}, $Z_b(10610)/Z_b(10650)$ \cite{Belle:2011aa}, $P_c(4380)/P_c(4450)$ \cite{Aaij:2015tga}, and so on.
Their unusual structure may contain more constituents, such as $\bar q q$ pairs or gluons.
With the extra $\bar q q$ pairs, multiquark configurations may form a compact structure with colored correlations, such as tetraquark and triquark, or a rather extended structure with color-singlet hadronic correlations, which is called the hadronic molecule \cite{Chen:2016qju,Liu:2013waa,Hosaka:2016pey}.

Since many new findings are seen near the threshold of hadronic decays, it is natural that the hadronic molecularlike structure develops if suitable attractive interactions between the hadrons are available.
The strength of the interaction between color-singlet hadrons should be weaker than that between colored objects of order $\Lambda_{QCD}$ -several hundred MeV.
A typical example of such hadronic molecules is an atomic nucleus whose binding energy is of order 10 or a few MeV.

For the study of hadronic molecules, the interaction between the hadrons is a crucial input.
Unfortunately, not much is known for the hadron interactions, which are relevant for the recent exotic hadrons.
For example, for $X(3872)$, regarded as $D \bar D^*$ molecule \cite{Liu:2008fh,Thomas:2008ja,Lee:2009hy,Li:2012cs,Sun:2012zzd,Zhao:2014gqa}, the realistic interaction between $D$ and $\bar{D}^*$ mesons is not well known partly because there is no experimental data.
Lattice QCD approaches should, in principle, be promising as the recent study for $Z_c(3900)$ \cite{Ikeda:2016zwx}. However, application to various systems is rather limited.
In such a situation, perhaps, the one-boson-exchange model is a reasonable theoretical approach.

According to the mass differences for the exchanged meson, the interactions from $\pi$, $\sigma$, $\rho$, and $\omega$ exchanges contribute in the long-range, intermediate-range, and short-range distances, respectively. Among them, the one pion exchange is the best known, as is important for the deuteron \cite{Tornqvist:1993vu,Tornqvist:1993ng} and the $X/Y/Z$ states \cite{Liu:2008fh,Thomas:2008ja,Lee:2009hy}. For the vector meson $\rho$, based on the local hidden gauge approach, it is also very essential in identifying the heavy molecular state \cite{Wu:2010jy}. In Ref. \cite{Karliner:2016ith}, the $\eta$ exchange is proposed to form heavy hadronic molecules. Soon after, the one-$\eta$-exchange model was adopted to investigate the interaction of $\Lambda_c\bar{D}_s^*/\Sigma_c^{(*)}\bar{D}_s^*/\Xi^{(\prime,*)}_c\bar{D}^*$ systems in Ref. \cite{Chen:2016ryt}. Numerical results indicate that the one-$\eta$ exchange can be helpful in binding the heavy molecular pentaquarks.

For the one-$\sigma$-exchange (OSE) and one-$\omega$-exchange (OOE) models, they have been always considered together with the other bosons $(\pi, \eta, \rho)$ in heavy molecular states, and their effect has been submerged by the effect of the one-$\pi$-exchange model. Thus, their importance is often overlooked. Therefore, the purpose of this paper is to study systematically the role of the OSE and OOE interactions between heavy hadrons.
The coupling strengths and form factors are estimated by using the quark model, where the sigma and omega mesons couple to light quarks in heavy hadrons.
Then we investigate if the intermediate- and short-range forces, due to the OSE and OOE models, can be strong enough to form heavy molecules, by varying model parameters within a reasonable range.

To elucidate the role of the $\sigma$ and $\omega$ mesons, we consider the systems, where $\pi, \eta, \rho$ meson exchanges are suppressed, by using the spin and isospin conservation. For instance, there is no coupling $\pi \Lambda_c \Lambda_c$ and $\pi D D$.  The pion couples rather in the transitions such as $\pi \Lambda_c \Sigma_c$ and $\pi DD^*$, which leads to the coupled channel problem.
In our present study, we focus exclusively on the $\sigma$ and $\omega$ mesons exchange by ignoring such coupled channels. Then, the $D\bar{D}$\footnote{The $\rho$ exchange can be also unsuppressed for the $D\bar{D}$ system. Here, it is considered to contrastively discuss the relation of the effective potentials from the OSE and OOE model.}, $\Lambda_cD$, and $\Lambda_c\bar{\Lambda}_c$\footnote{We also notice there are several former works on the $\Lambda_c\Lambda_c(\bar{\Lambda}_c)$ interactions \cite{Meguro:2011nr,Li:2012bt,Lee:2011rka}.} systems are the ones that we study in this paper. We then compare the properties of those systems where the $\sigma$ and $\omega$ mesons couple differently depending on the numbers of light quarks and antiquarks in the relevant hadrons.

This paper is organized as follows. After the Introduction, we derive the one-boson exchange (OBE) effective potentials in Sec. \ref{sec2}. In Sec. \ref{sec3}, we present the corresponding numerical results. Then, according to these conclusions, heavy quark-mass dependence is studied by varying it continuously in Sec. \ref{sec4}. The paper ends with a summery in Sec. \ref{sec5}.

\section{Interactions}\label{sec2}

\subsection{Lagrangians}

According to the heavy-quark symmetry, the OSE and OOE Lagrangians are constructed as
\begin{eqnarray}
\mathcal{L}_{DD\sigma/\omega} &=& -2g_{\sigma}DD^{\dag}\sigma+2g_{\omega} DD^{\dag}\bm{v}\cdot \bm{\omega},\label{DD}\\
\mathcal{L}_{\Lambda_c\Lambda_c\sigma/\omega} &=& -2g_{\sigma}'\bar{\Lambda}_c\Lambda_c\sigma-2g_{\omega}'\bar{\Lambda}_c\Lambda_c\bm{v}\cdot\bm{\omega}.\label{LL}
\end{eqnarray}
Here, $\bm{v}$ is the four velocity, which has the form of ${\bm{v}}=(1,\bf{0})$.

The coupling constants in Eqs. (\ref{DD}) and (\ref{LL}) will be determined in the quark model. Since the $\sigma$ and $\omega$ mesons couple dominantly to the light quarks, the relevant interaction Lagrangian for the light quarks $(q=u,d)$ with $\sigma/\omega$ can be expressed as
\begin{eqnarray}
\mathcal{L}_{qq\sigma/\omega} &=& -g_{\sigma}^{q}\bar{\psi}\sigma\psi-g_{\omega}^{q}\bar{\psi}\gamma^{\mu}\omega_{\mu}\psi.\label{quark}
\end{eqnarray}
Compared with the vertices of $D-D-\sigma/\omega$, $\Lambda_c-\Lambda_c-\sigma/\omega$, and $q-q-\sigma/\omega$, all the coupling constants in Eqs. (\ref{DD})$-$(\ref{quark}) can be related, i.e.,
\begin{eqnarray}\label{cc}
g_{\sigma} = g_{\sigma}'=g_{\sigma}^{q}, \quad\quad
g_{\omega} = g_{\omega}'=g_{\omega}^{q}.
\end{eqnarray}
In a $\sigma$ model \cite{Riska:1999fn}, the value of $g_{\sigma}^{q}$ is taken as $g_{\sigma}^{q}=3.65$. For the $\omega$ coupling $g_{\omega}^{q}$, it is of a little uncertainty; in the Nijmegen model, $g_{\omega}^{q}=3.45$, whereas it is equal to 5.28 in the Bonn model \cite{Rijken:1998yy}. In Ref. \cite{Riska:2000gd}, $g_{\omega}^{q}$ was roughly assumed to be 3.00. In the following calculation, all the possible choices will be employed.

According to the effective Lagrangians in Eqs. (\ref{DD}) and (\ref{LL}), all the relevant OBE scattering amplitudes can be collected in Table \ref{amplitude}.

\renewcommand\tabcolsep{0.1cm}
\renewcommand{\arraystretch}{2.2}
\begin{table}[htbp]
  \caption{Scattering amplitudes for all the investigated systems. Here, function $\mathcal{H}(\bm{q},m)$ is defined as $\mathcal{H}(\bm{q},m)=1/(\bm{q}^2+m^2)$.}\label{amplitude}
  \begin{tabular}{cl}\toprule[2pt]
  $h_1h_2\to h_3h_4$            &$\mathcal{M}(h_1h_2\to h_3h_4)$ \\\hline
  $DD\to DD$
                   &$4M_D^2\left[g_{\sigma}^2\mathcal{H}(\bm{q},m_{\sigma})-g_{\omega}^2\mathcal{H}(\bm{q},m_{\omega})\right]$\\
  $D\bar{D}\to D\bar{D}$
                   &$4M_D^2\left[g_{\sigma}^2\mathcal{H}(\bm{q},m_{\sigma})+g_{\omega}^2\mathcal{H}(\bm{q},m_{\omega})\right]$\\
  $\Lambda_c\bar{D}\to \Lambda_c\bar{D}$
                   &$8M_{D}M_{\Lambda_c}\chi_3^{\dag}\chi_1\left[g_{\sigma}g_{\sigma}'\mathcal{H}(\bm{q},m_{\sigma})-g_{\omega}g_{\omega}'\mathcal{H}(\bm{q},m_{\omega})\right]$\\
  $\Lambda_c{D}\to \Lambda_c{D}$
                   &$8M_{D}M_{\Lambda_c}\chi_3^{\dag}\chi_1\left[g_{\sigma}g_{\sigma}'\mathcal{H}(\bm{q},m_{\sigma})+g_{\omega}g_{\omega}'\mathcal{H}(\bm{q},m_{\omega})\right]$\\
  $\Lambda_c\Lambda_c\to\Lambda_c\Lambda_c$
       &$16M_{\Lambda_c}^2\chi_3^{\dag}\chi_4^{\dag}\chi_1\chi_2\left[g_{\sigma}g_{\sigma}'\mathcal{H}(\bm{q},m_{\sigma})-g_{\omega}g_{\omega}'\mathcal{H}(\bm{q},m_{\omega})\right]$\\
  $\Lambda_c\bar{\Lambda}_c\to\Lambda_c\bar{\Lambda}_c$
       &$16M_{\Lambda_c}^2\chi_3^{\dag}\chi_4^{\dag}\chi_1\chi_2\left[g_{\sigma}g_{\sigma}'\mathcal{H}(\bm{q},m_{\sigma})+g_{\omega}g_{\omega}'\mathcal{H}(\bm{q},m_{\omega})\right]$\\
  \bottomrule[2pt]
  \end{tabular}
\end{table}

Here, for the derivation of effective potentials of the $D\bar{D}$, $\Lambda_c\bar{D}$, and $\Lambda_c\bar{\Lambda}_c$ systems, the G-parity rule \cite{Klempt:2002ap} is adopted, which relates the scattering amplitudes between the processes $a+b\to c+d$ and $a+\bar{b}\to c+\bar{d}$ by exchanging one light meson.

With the help of the Breit approximation, a relation between the effective potentials in momentum space and the scattering amplitudes is obtained, i.e.,
\begin{eqnarray}
\mathcal{V}_{E}(\bm{q}) &=& -\frac{\mathcal{M}(h_1h_2\to h_3h_4)}{\sqrt{\prod_i2M_i\prod_f2M_f}}.
\end{eqnarray}
Here, $\mathcal{M}(h_1h_2\to h_3h_4)$ is defined as the scattering amplitude of the process $h_1h_2\to h_3h_4$. $M_i$ and $M_f$ are the masses of the initial states ($h_1$, $h_2$) and final states ($h_3$, $h_4$), respectively.

\subsection{Form factors}
The effective potential in the coordinate space $\mathcal{V}(r)$ is obtained by performing the Fourier transformation as
\begin{eqnarray}\label{fourier}
\mathcal{V}_{E}(\bm{r}) &=& \int\frac{d^3\bm{q}}{(2\pi)^3}e^{i\bm{q}\cdot\bm{r}}\mathcal{V}_{E}(\bm{q})\mathcal{F}^2(q^2).
\end{eqnarray}
In order to manipulate the off shell effect of the exchanged mesons $\sigma$ and $\omega$ and finite size effect of the interacting hadrons, we introduce a form factor $\mathcal{F}(q^2)$ at every vertex.

Generally, the form factor has the monopole, dipole, and exponential forms
\begin{eqnarray}
\mathcal{F}_M(q^2) &=& \frac{\Lambda^2-m^2}{\Lambda^2-q^2},\nonumber\\
\mathcal{F}_D(q^2) &=& \frac{\left(\Lambda^2-m^2\right)^2}{\left(\Lambda^2-q^2\right)^2},\nonumber\\
\mathcal{F}_E(q^2) &=& e^{\left(q^2-m^2\right)/\Lambda^2}.\nonumber
\end{eqnarray}
Here, $\Lambda$, $m$, and $q$ correspond to the cutoff, mass and momentum of the exchanged meson, respectively. These three kinds of form factors  are normalized at the on shell momentum of $q^2 = m^2$. In the low momentum limit, these form factors may be related to each other by redefining the cutoff parameter $\Lambda$ such that the first terms of the Taylor expansion in powers of $q^2/\Lambda^2$ coincide. In this way, low momentum phenomena of hadronic molecules do not depend very much on different choices of form factors.

The form factor can not be uniquely determined and various forms and cutoff $\Lambda$ are used phenomenologically.
However, an intuitive guideline for the choice of $\Lambda$ is done by relating it to the size of hadrons. In Refs. \cite{Tornqvist:1993vu,Tornqvist:1993ng}, $\Lambda$ is related to the root-mean-square radius of the source hadron to which the exchanged boson ($\sigma$ or $\omega$) couples. According to the previous experience of the deuteron, the cutoff $\Lambda$ in covariant-type monopole form factor is taken around 1 GeV. In the present qualitative study we use the same form factor both for meson and baryon vertices, because both of them contain light quarks and their spatial distributions are of order 1 fm or less.

\subsection{Effective potentials}\label{sub3}

In this subsection, we adopt the monopole form factor $\mathcal{F}_M(q^2)$, and the resulting effective potentials for the investigated systems are collected in Table \ref{tot}.
\renewcommand\tabcolsep{0.2cm}
\renewcommand{\arraystretch}{1.8}
\begin{table}[!htbp]
  \caption{Effective potentials for the investigated systems. The function $Y(\Lambda,m,r)$ is defined as $Y(\Lambda,m,{r}) = (e^{-mr}-e^{-\Lambda r})/4\pi r-(\Lambda^2-m^2)e^{-\Lambda r}/{8\pi \Lambda}$.}\label{tot}
  \begin{tabular}{ccl}\toprule[2pt]
  Systems      &Quarks    &$\mathcal{V}({r})$ \\\hline
  $DD$          &$(c\bar{q})(c\bar{q})$           &$-g_{\sigma}^2Y(\Lambda,m_{\sigma},r)+g_{\omega}^2Y(\Lambda,m_{\omega},r)$\\
     $D\bar{D}$          &$(c\bar{q})(\bar{c}{q})$           &$-g_{\sigma}^2Y(\Lambda,m_{\sigma},r)-g_{\omega}^2Y(\Lambda,m_{\omega},r)$\\
  $\Lambda_c\bar{D}$     &$(cqq)(\bar{c}{q})$         &$-2g_{\sigma}g_{\sigma}'Y(\Lambda,m_{\sigma},r)+2g_{\omega}g_{\omega}'Y(\Lambda,m_{\omega},r)$\\
     $\Lambda_c{D}$     &$(cqq)(c\bar{q})$         &$-2g_{\sigma}g_{\sigma}'Y(\Lambda,m_{\sigma},r)-2g_{\omega}g_{\omega}'Y(\Lambda,m_{\omega},r)$\\
  $\Lambda_c\Lambda_c$    &$(cqq)(cqq)$          &$-4g_{\sigma}^{'2}Y(\Lambda,m_{\sigma},r)+4g_{\omega}^{'2}Y(\Lambda,m_{\omega},r)$\\
      $\Lambda_c\bar{\Lambda}_c$    &$(cqq)(\bar{c}\bar{q}\bar{q})$          &$-4g_{\sigma}^{'2}Y(\Lambda,m_{\sigma},r)-4g_{\omega}^{'2}Y(\Lambda,m_{\omega},r)$\\
  \bottomrule[2pt]
  \end{tabular}
\end{table}

In Table \ref{tot}, we can find that the interactions from the OSE model are always attractive for these investigated systems. This is a general consequence of the scalar meson exchange with a momentum independent coupling constant, as briefly explained in the next section. The depth of the OSE effective potentials depends on the number of the light quarks and/or antiquark combinations $(q-q, q-\bar{q}, \bar{q}-\bar{q})$, where the light quark or antiquark is reserved in different hadrons of the hadron-hadron systems, respectively. For example, according to the quark configurations as shown in the second column of the Table \ref{tot}, the light $\bar{q}-\bar{q}$ combination for the $DD$ system is one, and there is only one $q-\bar{q}$ combination in the $D\bar{D}$ system.

Since $g_{\sigma}=g_{\sigma}'$ as estimated in the quark model (\ref{cc}), a simple relation between the OSE effective potentials and the light-quark and/or antiquark combination numbers can be summarized as
\begin{eqnarray}
\mathcal{V}_{\sigma}(x_{qq/\bar{q}\bar{q}},y_{q\bar{q}}) &=& -(x_{qq/\bar{q}\bar{q}}+y_{q\bar{q}})g_{\sigma}^2Y(\Lambda,m_{\sigma},r),\label{vs}
\end{eqnarray}
where $x_{qq/\bar{q}\bar{q}}$ and $y_{q\bar{q}}$ correspond to the numbers of $qq/\bar{q}\bar{q}$ and $q\bar{q}$ $(q=u,d)$ combinations, respectively. For the OOE effective potentials, a similar relation can be also written as 
\begin{eqnarray}
\mathcal{V}_{\omega}(x_{qq/\bar{q}\bar{q}},y_{q\bar{q}}) &=& (x_{qq/\bar{q}\bar{q}}-y_{q\bar{q}})g_{\omega}^2Y(\Lambda,m_{\omega},r).\label{vo}
\end{eqnarray}
Here, we note that the sign of the OOE changes according to the charge conjugation symmetry. For example, the OOE force is repulsive for the $q-q$ and $\bar{q}-\bar{q}$ combinations, while reversed for the system with $q-\bar{q}$ combination.

For example, for the $\Lambda_cD$ system, there are two $q-\bar q$ combinations, thus $y_{q\bar{q}}=2$, and its potential from $\sigma$ and $\omega$ exchanges is
\begin{eqnarray}
\mathcal{V}_{\Lambda_cD}(r) &=& -2g_{\sigma}^2Y(\Lambda,m_{\sigma},r)-2g_{\omega}^2Y(\Lambda,m_{\omega},r).
\end{eqnarray}

\begin{figure}[!htbp]
\center
\includegraphics[width=3.4in]{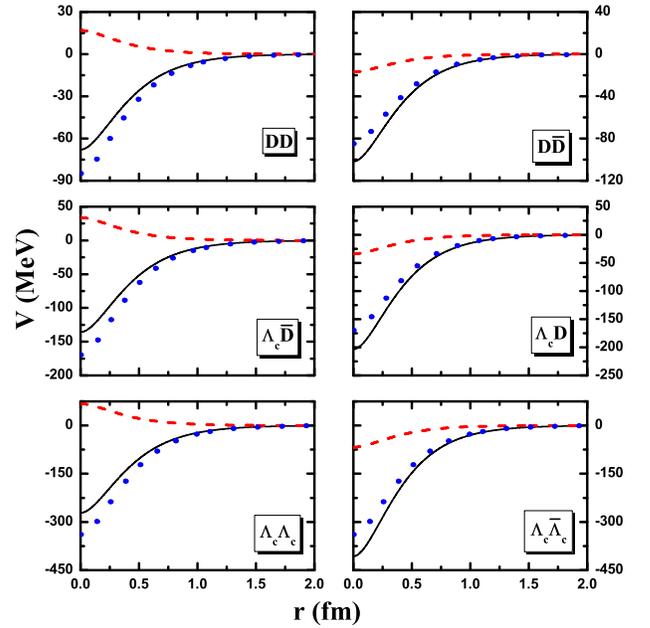}
\caption{(color online) Effective potentials for the $DD(\bar{D})$, $\Lambda_cD(\bar{D})$, and $\Lambda_c\Lambda_c(\bar{\Lambda}_c)$ systems with cutoff value $\Lambda=1.00$ GeV, $g_{\sigma}^{q}=3.65$, and $g_{\omega}^{q}=3.00$. Here, the dotted, dashed, and solid lines correspond to the OSE, OOE, and total effective potentials, respectively.}\label{ps}
\end{figure}

In Fig. \ref{ps}, we present the resulting potential as functions of the distance $r$,
where the total potential is shown by the solid line, OSE by dotted lines and OOE by dashed lines.

Here, the OSE and the OOE forces are of typical character of intermediate- and short-range force, and therefore they are suppressed  when the radius $r$ reaches 1 fm and larger. Since the force from the OSE model is the dominant, the total effective potentials for all the investigated systems are all attractive.

To summarize shortly, the OSE model can always provide an attractive force. However, the OOE force is repulsive for the system including the same light quarks or antiquarks in its components of the investigated systems. The interaction strength from the OSE and OOE models depends on the light-quark combination numbers.

\section{Numerical results}\label{sec3}

In this section we discuss the role of the OSE and OOE interaction for the systems of $\Lambda_c D(\bar D)$ and $\Lambda_c \Lambda_c (\bar \Lambda_c)$ by solving the Schr\"{o}dinger equation for them
\begin{eqnarray}
-\frac{1}{2M}\nabla^2\psi(r)+V(r)\psi(r)=E\psi(r),
\end{eqnarray}
where $\nabla^2=\frac{1}{r^2}\frac{\partial}{\partial r}r^2\frac{\partial}{\partial r}$, and $M=m_1m_2/(m_1+m_2)$ is the reduced mass for the investigated system composed by particle 1 and particle 2. The parameters we use are summarized in Table \ref{parameter}.
\renewcommand\tabcolsep{0.16cm}
\renewcommand{\arraystretch}{1.8}
\begin{table}[!htbp]
  \caption{Parameters adopted in this work \cite{Olive:2016xmw}.}\label{parameter}
  \begin{tabular}{ccc|ccc}\toprule[2pt]
  Hadron     &$I(J^P)$     &Mass (MeV)    &Hadron     &$I(J^P)$     &Mass (MeV) \\\hline
  $K$        &$\frac{1}{2}(0^-)$    &495.64       &$\Lambda$       &$0(\frac{1}{2}^+)$     &1115.683\\
  $D$        &$\frac{1}{2}(0^-)$    &1867.24      &$\Lambda_c$     &$0(\frac{1}{2}^+)$     &2286.46\\
  $B$        &$\frac{1}{2}(0^-)$    &5279.42      &$\Lambda_b$     &$0(\frac{1}{2}^+)$     &5619.4\\
  $\sigma$   &$0(0^+)$              &600          &$\omega$        &$0(1^-)$               &782.65\\
  \bottomrule[2pt]
  \end{tabular}
\end{table}

\subsection{{{Solutions with covariant-type monopole form factor $\mathcal{F}_M(q^2) = (\Lambda^2-m^2)/(\Lambda^2-q^2)$}}}

We summarize the properties of S-wave bound states when they exist and the binding energy and root-mean-square radii ($r_{RMS}$) for the S-wave $\Lambda_c D(\bar{D})$ and $\Lambda_c\Lambda_c(\bar{\Lambda}_c)$ systems in Table \ref{num1}.
For the coupling constant $g_{\omega}^{q}$, we use three values (3.00 for Case I, 3.45 for II, and 5.28 for III) corresponding to the $g_{\omega}^{q}$ coupling constants of Ref. \cite{Riska:2000gd}, of the Nijmegen model, and of the Bonn model \cite{Rijken:1998yy}, respectively.
%The cutoff parameters are chosen around 1 GeV \cite{Tornqvist:1993vu,Tornqvist:1993ng}.
{{In Table \ref{num1}, the cutoff parameters are chosen as 1, 1.1, and 1.2 GeV. These are the typical values for the form factor $\mathcal{F}_M(q^2)$ \cite{Tornqvist:1993vu,Tornqvist:1993ng}.
In fact, depending on detailed values of $\Lambda$ and on channels, bound states may or may not appear. In this way, we discuss whether bound states appear or not and study the role of the $\sigma$ and $\omega$ meson exchanges.}}

%These are the typical values for the systems to allow loosely bound molecular states with binding energies of order at most several tens MeV.
%Within these ranges, we discuss whether bound states appear or not and study the role of the $\sigma$ and $\omega$ meson exchanges.

\renewcommand\tabcolsep{0.3cm}
\renewcommand{\arraystretch}{1.8}
\begin{table*}[!htbp]
  \caption{Bound solutions for the S-wave $\Lambda_cD(\bar{D})$ and $\Lambda_c\Lambda_c(\bar{\Lambda}_c)$ systems. Here, the monopole form factor $\mathcal{F}_M(q^2)=(\Lambda^2-m^2)/(\Lambda^2-q^2)$ is adopted. The units for cutoff $\Lambda$, binding energy $E$ and root-mean-square radius $r_{RMS}$ are GeV, MeV, and fm, respectively. Cases I, II, and III correspond to the numerical results by adopting the coupling constant $g_{\omega}^{q}$ taken with the value $g_{\omega}^{q}=3.00$ in Ref. \cite{Riska:2000gd}, 3.45 in the Nijmegen model, and 5.28 in the Bonn model \cite{Rijken:1998yy}, respectively. The notation $\ldots$ stands for no bound solutions.}\label{num1}
  \begin{tabular}{c|ccccccc||ccccccc}\toprule[2pt]
     \multirow{2}*{\,Cases\,}     &\multicolumn{3}{c}{$\Lambda_c\bar D$}    &    &\multicolumn{3}{c||}{$\Lambda_cD$}     &\multicolumn{3}{c}{$\Lambda_c\Lambda_c$}    &   &\multicolumn{3}{c}{$\Lambda_c\bar{\Lambda}_c$}\\\cline{2-15}

            &$\Lambda$     &$E$     &$r_{RMS}$          &           &$\Lambda$        &$E$     &$r_{RMS}$
            &$\Lambda$     &$E$     &$r_{RMS}$          &            &$\Lambda$        &$E$     &$r_{RMS}$\\\hline
     I
          &1.00        &\ldots        &\ldots       &    &1.00             &-1.61         &2.79
          &1.00      &-17.47    &1.06             &      &1.00             &-50.49        &0.73\\

         &1.10        &-0.04        &6.27        &   &1.10             &-11.09         &1.26
         &1.10        &-31.70      &0.85             &   &1.10         &-110.39        &0.55\\

         &1.20        &-0.90        &3.53       &   &1.20             &-27.66         &0.89
         &1.20        &-46.84       &0.74             &   &1.20       &-187.32        &0.46\\\hline

  II     &1.00        &\ldots        &\ldots         &      &1.00             &-2.41         &2.35
         &1.00      &-13.52    &1.16         &       &1.00             &-56.99        &0.70\\

         &1.10        &\ldots        &\ldots          &     &1.10             &-14.40         &1.14
         &1.10        &-22.74        &0.96        &      &1.10                &-126.02        &0.53\\

         &1.20        &\ldots        &\ldots         &      &1.20             &-34.85         &0.82
         &1.20        &-31.46        &0.85              &      &1.20          &-215.32        &0.44\\\hline

  III     &1.00        &\ldots        &\ldots   &     &1.00             &-9.32         &1.36
          &1.00      &-0.60    &4.01              &     &1.00             &-97.49        &0.58\\

          &1.10        &\ldots        &\ldots   &     &1.10             &-38.72         &0.80
          &1.10        &\ldots        &\ldots               &    &1.10  &-222.80        &0.44\\

          &1.20        &\ldots        &\ldots    &    &1.20             &-85.10         &0.60
          &1.20      &\ldots        &\ldots               &    &1.20    &-387.71        &0.36\\
  \bottomrule[2pt]
  \end{tabular}
\end{table*}

Before discussing details of Table \ref{num1}, we make general remarks for boson exchange potentials.

\begin{itemize}

\item
The $\sigma$ meson exchange provides attractive interaction.
This is understood using a second-order perturbation theory for the one boson-exchange; the intermediate three particle state with $\sigma$ meson has a virtual energy that is larger than the initial (or final) energy of the two particles.
Moreover, due to the positive charge conjugation of the Lorentz scalar charge that the $\sigma$ meson couples to, the signs of the couplings for both quark and antiquark are the same.
This explains the universally attractive nature of the $\sigma$ meson exchange.

\item
In comparison with the $\sigma$ exchange, the $\omega$ meson
couples to the baryonic charge which flips its sign for quark and antiquark.
This provides a repulsive interaction between quarks and attractive interaction between the quark and antiquark.

\item
The role of $\Lambda$ is to suppress the interaction strength for larger momentum transfer and thus effectively reduce the strength of the interaction for bound states.
As we will see, the results depend very much on the choice of the form factor.

\end{itemize}

For $\Lambda_c \bar D$, the interaction is the sum of attractive OSE and repulsive OOE, with the total is some attractive.
As $\Lambda$ is increased, the OSE becomes more prominent, and a bound state appears for $\Lambda > 1.1$ GeV for case I.
For cases II and III, because of slightly stronger $\omega$ exchange repulsion, we do not find any bound states.
These are the results for S waves.
For higher partial waves, due to the repulsive centrifugal force, $l(l+1)/2Mr^2$,
it is less likely to have bound states.
{{Thus, we may conclude that in our model with a reasonable $\Lambda\sim$ 1 GeV, hidden-charm molecular pentaquarks made up by $\Lambda_c \bar D$ are not likely to exist.
Indeed, if we increase $\Lambda$ larger than 1 GeV when more attraction is expected, we do not yet find bound states or do, at most, very weakly bound states only for case I.}}
Experimentally, our conclusion for the $\Lambda_c\bar{D}$ system is consistent with the current results of LHCb \cite{Aaij:2015tga}, where no obvious evidence of possible partners of $P_c(4380)$ and $P_c(4450)$ has been reported, in the region close to the mass of $\Lambda_c\bar{D}$.

As compared to the $\Lambda_c\bar{D}$ system, the OOE force for the $\Lambda_cD$ system is attractive, as explained above. Together with the attractive OSE force, the net attractive force for the $\Lambda_cD$ turns out to be strong enough to accommodate bound states. As shown in Table \ref{num1}, for the cutoff $\Lambda\sim 1$ GeV, we find a shallow bound state with a binding energy around several MeV.
Therefore, this channel may provide a good candidate of a loosely bound molecular state of the $\Lambda_cD$ system with $|{}^2S_{\frac{1}{2}}\rangle$. Since $D$ and $\Lambda_c$ are the lowest ground hadrons of the charmed mesons and baryons, its possible strong decay channel should be rather limited, like $\Xi_{cc}(\frac{1}{2}^+)+\pi/\eta$.

For the heavy baryon systems $\Lambda_c\Lambda_c(\bar{\Lambda}_c)$, since one more
light quark (antiquark) is in the baryon $\Lambda_c(\bar \Lambda_c)$,
the interaction strength becomes two times stronger than
that in the $\Lambda_c\bar{D}$ and $\Lambda_cD$ systems.
Therefore, as shown in Table \ref{num1},
more bound state solutions have been found both for $\Lambda_c\Lambda_c$ and $\Lambda_c\bar{\Lambda}_c$ systems than for $\Lambda_c\bar{D}$ and $\Lambda_cD$ systems.
With the same cutoff input, their binding energies reach several tens MeVs. Thus, they can be also possible molecular candidates. For their decay behaviors,
the $\Xi_{cc}(\frac{1}{2}^+)N$ can be the only strong decay channel for the S-wave $\Lambda_c\Lambda_c$ bound state. The decay processes will be much more complicated for the S-wave $\Lambda_c\bar{\Lambda}_c$ molecular state, as they include open-charm and hidden-charm channels, like $\chi_{c0}+\pi\pi$, $D\bar{D}_1+\pi$, and so on.

\subsection{{{Solutions with noncovariant-type monopole form factor $\mathcal{F}_M(q^2) = \Lambda^2/(\Lambda^2-q^2)$}}}

So far, we have employed a covariant monopole form factor and discussed the role of OSE and OOE potentials, with some predictions for molecular candidates, the S-wave $\Lambda_cD$, $\Lambda_c\Lambda_c$, and $\Lambda_c\bar{\Lambda}_c$.
In order to further see our discussions, in the following, we attempt to use a three-momentum form factor of the form of $\mathcal{F}(q^2)=\Lambda^2/(\Lambda^2-q^2)$, which is often adopted in nuclear physics.

In the nonrelativistic kinematics, the energy transfer is neglected, and so this condition reduces to the condition of vanishing three momentum.
%This form factor is related to the monopole form factor by renormalizing the coupling constant,
%\begin{eqnarray}
%\mathcal{F}(q^2)&=&\left(1-\frac{m^2}{\Lambda^2}\right)\mathcal{F}_M(q^2).\nonumber
%\end{eqnarray}
In fact, the difference of this form factor from the monopole form factor is absorbed into the redefinition of the coupling constants as Eqs. (\ref{redefine1}) and (\ref{redefine2})
\begin{eqnarray}
f_{\sigma} &=& f'_{\sigma} = \left(1-\frac{m_{\sigma}^2}{\Lambda^2}\right)g_{\sigma} = \left(1-\frac{m_{\sigma}^2}{\Lambda^2}\right)g'_{\sigma},\label{redefine1}\\
f_{\omega} &=& f'_{\omega} = \left(1-\frac{m_{\omega}^2}{\Lambda^2}\right)g_{\omega} = \left(1-\frac{m_{\omega}^2}{\Lambda^2}\right)g'_{\omega}.\label{redefine2}
\end{eqnarray}
If we use the same coupling constants and cutoff $\Lambda$, the interaction strengths are larger when the three-dimensional form factor is employed.
Therefore, to obtain loosely bound molecular states, we need to use smaller cutoff $\Lambda$ when the coupling constants are kept unchanged.
This is the reason that we show the results in Table \ref{num2} with smaller cutoff $\Lambda$.

In order to determine the value of cutoff in noncovariant-type monopole form factor, here, we recall the relation,
\begin{eqnarray}
\langle r^2 \rangle &=& -6\left.\frac{\partial\mathcal{F}(q^2)}{\partial q^2}\right|_{q^2\to 0} \approx \frac{6}{\Lambda^2}. \label{cutoff}
\end{eqnarray}
If the form factor $\mathcal{F}(q^2)$ is introduced, in practice, the resulting cutoff parameter for $\mathcal{F}$ turns out to be around 0.5 GeV as we discussed around Eq. (7),
consistent with typical hadronic size. The results are shown in Table \ref{num2} for $\Lambda\sim$ 0.4, 0.5, and 0.6 GeV.

\renewcommand\tabcolsep{0.3cm}
\renewcommand{\arraystretch}{1.8}
\begin{table*}[!htbp]
  \caption{Bound solutions for the S-wave $\Lambda_cD(\bar{D})$ and $\Lambda_c\Lambda_c(\bar{\Lambda}_c)$ systems. Here, the form factor $\mathcal{F}(q^2)=\Lambda^2/(\Lambda^2-q^2)$ is adopted. The units for cutoff $\Lambda$, binding energy $E$, and root-mean-square radius $r_{RMS}$ are GeV, MeV, and femtometer, respectively. Cases I, II, and III correspond to the numerical results by adopted the coupling constant $g_{\omega}^{q}$ taken the value $g_{\omega}^{q}=3.00$ in Ref. \cite{Riska:2000gd}, 3.45 in the Nijmegen model, and 5.28 in the Bonn model \cite{Rijken:1998yy}, respectively. The notation $\ldots$ stands for no bound solutions.}\label{num2}
  \begin{tabular}{c|ccccccc||ccccccc}\toprule[2pt]
     \multirow{2}*{\,Cases\,}     &\multicolumn{3}{c}{$\Lambda_c\bar D$}    &    &\multicolumn{3}{c||}{$\Lambda_cD$}     &\multicolumn{3}{c}{$\Lambda_c\Lambda_c$}    &   &\multicolumn{3}{c}{$\Lambda_c\bar{\Lambda}_c$}\\\cline{2-15}

            &$\Lambda$     &$E$     &$r_{RMS}$          &           &$\Lambda$        &$E$     &$r_{RMS}$
            &$\Lambda$     &$E$     &$r_{RMS}$          &            &$\Lambda$        &$E$     &$r_{RMS}$\\\hline
     I
          &0.40        &\ldots        &\ldots             &    &0.40             &-6.94         &1.81
          &0.40        &-4.08         &2.14               &    &0.40             &-45.87        &0.95 \\

          &0.50        &\ldots        &\ldots        &   &0.50             &-16.55         &1.28
          &0.50        &-9.51         &1.51          &   &0.50             &-88.07         &0.73\\

          &0.60        &-0.15        &5.79       &   &0.60        &-30.66         &1.00
          &0.60        &-17.20       &1.20       &   &0.60        &-144.37        &0.60\\\hline

  II      &0.40        &\ldots        &\ldots             &    &0.40             &-9.38         &1.63
          &0.40        &-1.14         &3.45               &    &0.40             &-54.98        &0.90\\

          &0.50        &\ldots        &\ldots             &    &0.50             &-21.48        &1.17
          &0.50        &-3.31         &2.23               &    &0.50             &-104.66       &0.69\\

          &0.60        &\ldots        &\ldots             &   &0.60        &-38.96        &0.92
          &0.60        &-6.57         &1.69               &   &0.60        &-170.72       &0.56\\\hline

  III     &0.40        &\ldots        &\ldots             &    &0.40             &-26.97        &1.15
          &0.40        &\ldots        &\ldots             &    &0.40             &-112.02       &0.72\\

          &0.50        &\ldots        &\ldots             &    &0.50             &-55.46        &0.86
          &0.50        &\ldots        &\ldots             &    &0.50             &-207.12       &0.56\\

          &0.60        &\ldots        &\ldots             &    &0.60        &-94.71        &0.69
          &0.60        &\ldots        &\ldots             &    &0.60        &-331.99       &0.46\\
  \bottomrule[2pt]
  \end{tabular}
\end{table*}

Compared with the numerical results in Table \ref{num1}, one can find that, if we take a value of $\Lambda=0.5$ GeV, which is estimated by Eq. (\ref{cutoff}), the results in Table \ref{num2} are very similar to those in Table \ref{num1}. Having these results together with those of different form factors, we find that the intermediate-range and short-range force from OSE and OOE models provides a strong attraction to generate bound states.
%In fact, we also notice that the $\Lambda_c\Lambda_c(\bar{\Lambda}_c)$ systems have been studied \cite{Meguro:2011nr,Li:2012bt,Lee:2011rka}.

{{Finally, let us give a brief conclusion, where we show the results with the two form factors $\mathcal{F}_M$ ($\Lambda\sim$ 1 GeV) in Table \ref{num1}
and those with $\mathcal{F}$($\Lambda\sim$ 0.5 GeV) in Table \ref{num2}. To be seen shortly, these results are
qualitatively similar but have some differences quantitatively. The latter indicates uncertainties of the present model calculations. Nevertheless, we can predict several possible candidates for molecular states, S-wave $\Lambda_cD$, $\Lambda_c\Lambda_c$, and $\Lambda_c\bar{\Lambda}_c$ states.}}

\section{Extension}\label{sec4}

\subsection{Mass dependence}

In addition to the effective potentials, the mass in the kinetic term is another important input for the discussion of bound states. In fact, in the heavy-quark limit, $(M\to\infty)$ as the kinetic energy $p^2/2M$ vanishes, hadrons will be more easily bound. In the following, we study the reduced mass dependence of the molecular systems.

\begin{figure}[!htbp]
\center
\includegraphics[width=3.4in]{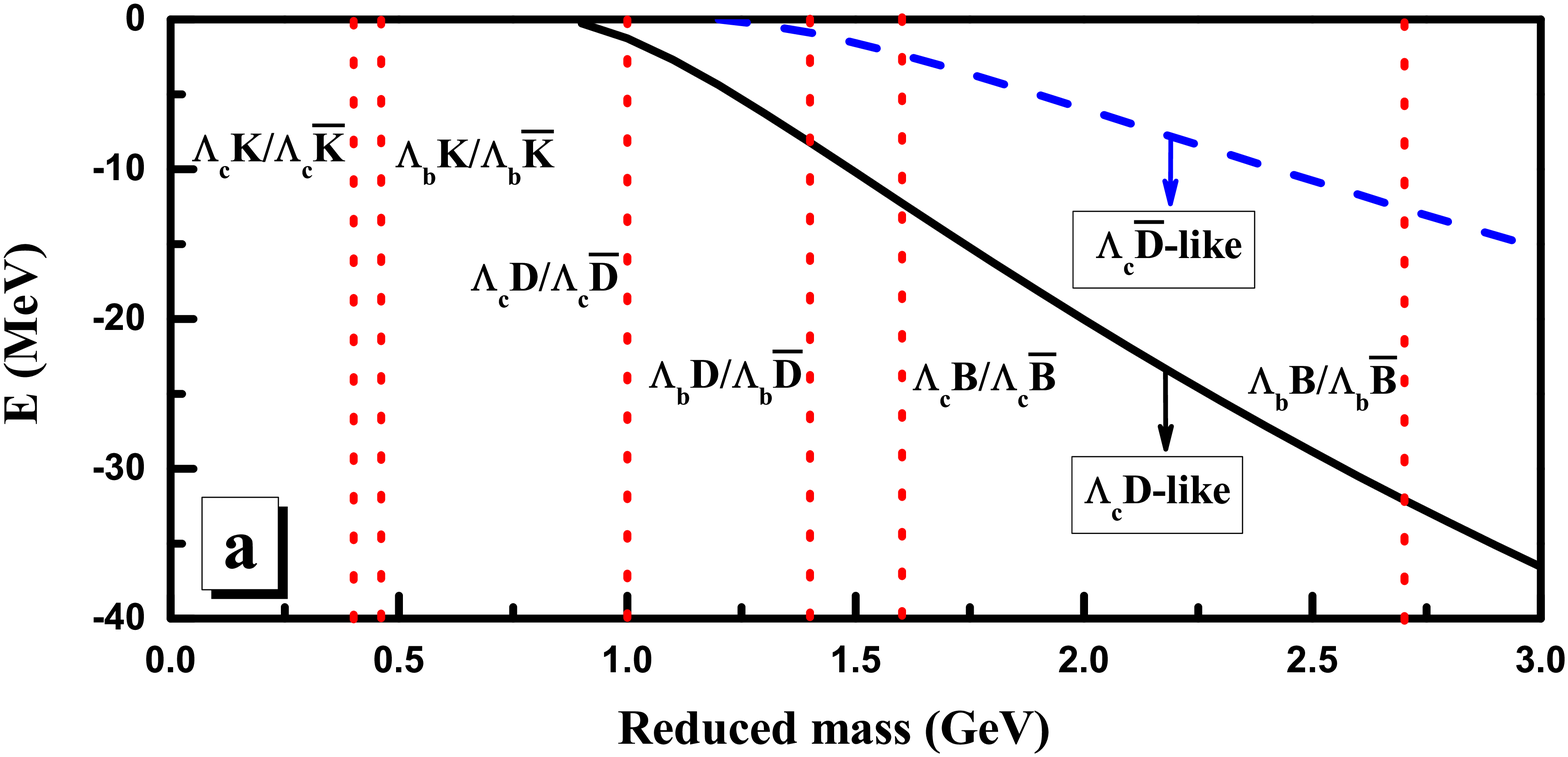}
\includegraphics[width=3.4in]{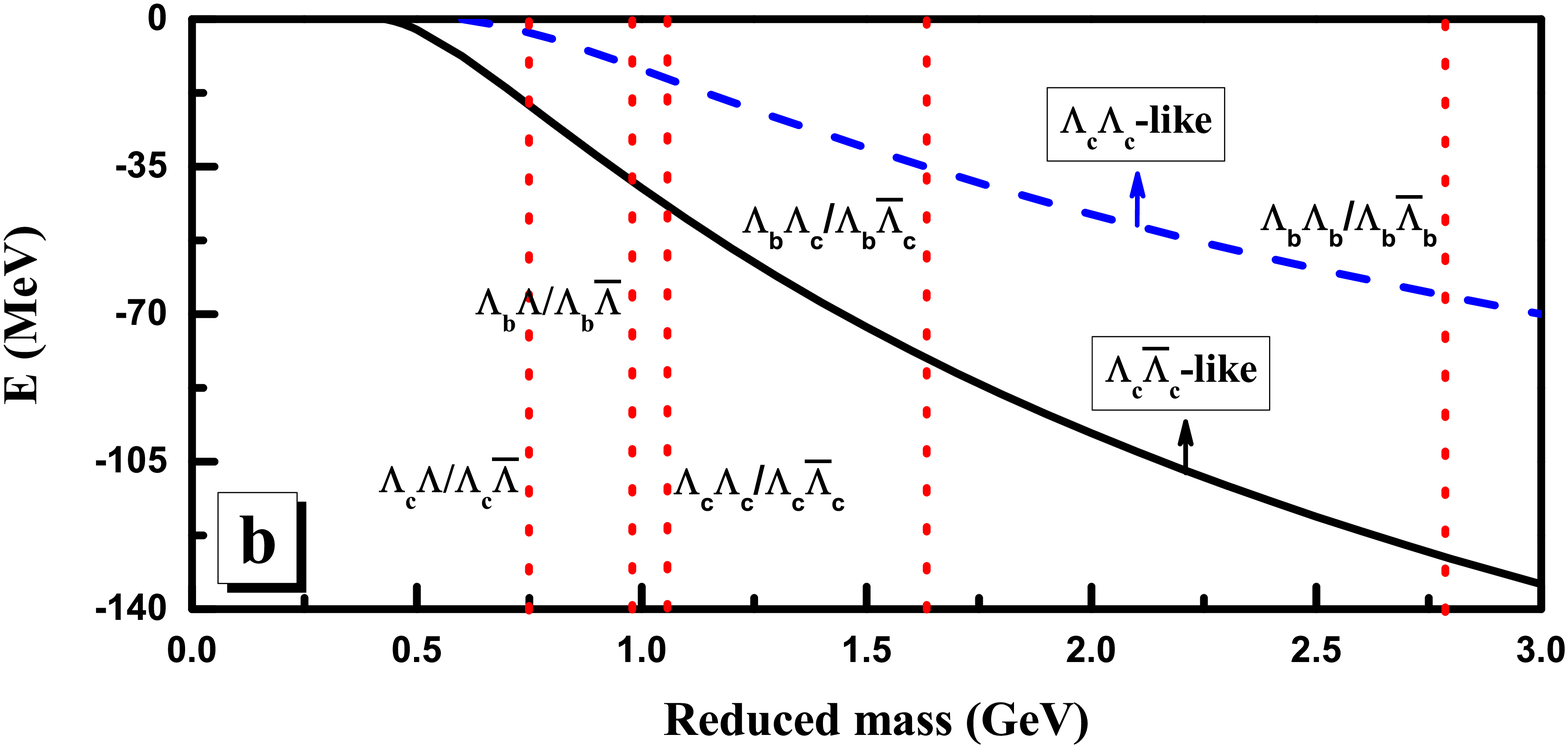}
\caption{ Binding energies as functions of the reduced mass. Here, $\Lambda=1.00$ GeV, $g_{\sigma}^{q}=3.65$, and $g_{\omega}^{q}=3.00$. The dotted lines stand for the reduced masses for several hadron-hadron systems as indicated. }\label{reduce}
\end{figure}

The upper panel of Fig. \ref{reduce} shows the binding energies of the $\Lambda_c \bar D-$like (dashed line) and $\Lambda_c D-$like (solid line) states, where the reduced mass of the two hadrons is varied as in the horizontal axis.
The vertical dotted lines correspond to the reduced masses of the two hadrons, as indicated in the figure.
The solid line stands for the binding energies of the $\Lambda_c D-$like state; it starts to appear when the reduced mass becomes larger than $\sim$ 0.75 GeV, and as expected, the binding energy increases as the reduced mass is increased.
For the $\Lambda_c \bar D-$like state, the OOE potential is repulsive, resulting in less attractive potential than for the $\Lambda_c D-$like state.
Thus, the system allows weaker binding as the dashed line shows.
The lower panel shows similar results for the $\Lambda_c \Lambda_c-$ and $\Lambda_c \bar \Lambda_c-$like states.
Because these systems have a larger attraction as proportional to the number of the light quarks as compared with the $\Lambda_c D$ and $\Lambda_c \bar D$ ones, larger binding energies are obtained.

When the reduced mass is sufficiently heavy, as in the charm and
bottom regions but not in the strange regions, binding energies for $\Lambda_c D$ and $\Lambda_c \bar D$ systems reach several to several tens MeV. Thus, the heavy flavors of charm and bottom are important in stabilizing pentaquark hadronic molecules.
For the $\Lambda_c\Lambda_c(\bar{\Lambda}_c)$-like systems, with stronger OSE and OOE interactions, more bound solutions are obtained even in the strangelike section.
Therefore, the heavy dibaryon molecules can be more stable than the heavy pentaquark molecules.
These results suggest that searching for the heavy dibaryon molecules is very promising in experiments.

\subsection{$\Lambda_cN$ and $\Lambda_c\bar{N}$ systems}

In this subsection, let us study $\Lambda_c N (\bar N)$ systems. In fact, the $\Lambda_c N (\bar N)$ interactions have been investigated \cite{Dover:1977jw,Liu:2011xc,Meng:2017udf}. In particular, a very shallow bound state was found for the S-wave $\Lambda_cN$ system \cite{Maeda:2015hxa}. According to Eqs. (\ref{vs}) and (\ref{vo}), the total effective potentials for $\Lambda_cN$, and $\Lambda_c\bar{N}$ systems are written as
\begin{eqnarray}
\mathcal{V}_{\Lambda_cN}(r) &=& -6g_{\sigma}^2Y(\Lambda,m_{\sigma},r)+6g_{\omega}^2Y(\Lambda,m_{\omega},r),\\
\mathcal{V}_{\Lambda_c\bar{N}}(r) &=& -6g_{\sigma}^2Y(\Lambda,m_{\sigma},r)-6g_{\omega}^2Y(\Lambda,m_{\omega},r).
\end{eqnarray}

The bound solutions for the S-wave $\Lambda_{c,b}N(\bar{N})$ systems are summarized in Table \ref{num3}. When we take the cutoff around 1 GeV, their binding energy reaches a few to several tens MeV, and their root-mean-square radii are around 1 fm. This means that they also can be possible molecular candidates. For $\Lambda_cN$ and $\Lambda_bN$ states, if they form bound states, they are stable under the strong interaction, while the $\Lambda_c\bar{N}$ and $\Lambda_b\bar{N}$ states can decay to charmed/antibottomed meson and the light mesons, like $D\pi\pi$ and $\bar{B}\pi\pi$.

\renewcommand\tabcolsep{0.3cm}
\renewcommand{\arraystretch}{1.4}
\begin{table}[!htbp]
  \caption{Bound solutions for the S-wave $\Lambda_cN$, $\Lambda_c\bar{N}$, $\Lambda_bN$, and $\Lambda_b\bar{N}$ systems. Here, the parameters are taken as $g_{\sigma}^{q}=3.65$, $g_{\omega}^{q}=3.00$. The units for cutoff $\Lambda$, binding energy $E$, and root-mean-square radius $r_{RMS}$ are GeV, MeV, and femtometer, respectively.}\label{num3}
  \begin{tabular}{c|cccccc}\toprule[2pt]
        &\multicolumn{2}{c}{$\Lambda_cN$}    &    &\multicolumn{2}{c}{$\Lambda_c\bar{N}$}  \\\hline
     $\Lambda$      &0.9        &1.0        &&0.9        &1.0  \\
     $E$            &-4.28      &-17.84     &&-11.92     &-59.97 \\
     $r_{RMS}$       &2.25       &1.28       &&1.50       &0.82\\ \midrule[1pt]
      &\multicolumn{2}{c}{$\Lambda_bN$}    &    &\multicolumn{2}{c}{$\Lambda_b\bar{N}$}  \\\hline
     $\Lambda$      &0.9        &1.0        &&0.9        &1.0\\
     $E$            &-10.38      &-29.84      &&-21.57      &-82.28\\
     $r_{RMS}$      &1.47        &0.99        &&1.12        &0.70\\
  \bottomrule[2pt]
  \end{tabular}
\end{table}

\section{Conclusion and discussion}\label{sec5}

Stimulated by the observation of $X/Y/Z/P_c$ states near threshold, the study of the hadronic molecular picture becomes more and more essential. For the study of molecular states, it is essentially important to describe the interaction between the hadrons of the molecular system. For this purpose, currently, the most available approach is the one-boson-exchange model based on the knowledge of light quark and boson interactions, which is applied to the system of open heavy hadrons containing light quarks. In this work, we systematically study the properties of the interaction from the one-$\sigma/\omega$-exchange model. The $\Lambda_cD(\bar{D})$, $\Lambda_c\Lambda_c(\bar{\Lambda}_c)$ systems have been taken into consideration. Meanwhile, all the parameters are estimated by the quark model.

In fact, the intermediate- and short-range interactions between hadrons are from a many pions exchange process. Here, $\sigma$ and $\omega$ exchanges are adopted to approximately replace two and three pion exchanges interactions, respectively. Compared to a pion, $\sigma$ meson is of uncertain mass and wide width, which affects the strength of the $\sigma-$exchange interaction. In the limit of small momentum transfer, the effective potential from $\sigma$ exchange is proportional to the term of $g_{\sigma}^2/m_{\sigma}^2$. Therefore, the uncertainty in the mass and wide width may be absorbed into the redefinition of the coupling constant. {{In nuclear physics, the mass of $\sigma$ is often taken as a fixed input parameter to fit the phase shift of nucleon-nucleon interaction \cite{Machleidt:1987hj,Backman:1984sx}.}}

By working out suitable coupling constants and form factors, we find that in many cases the sum of OSE and OOE models become attractive, where the OSE plays the dominant role. The OSE force is always attractive and the dominant. Whereas, for the OOE force, it is repulsive when there exists $q-q\,(q=u,d)$ or $\bar{q}-\bar{q}$ combinations in the two hadrons system. The OSE and OOE interaction strength depends on the number of $q-q$, $\bar{q}-\bar{q}$, and $q-\bar{q}$ combinations. With reasonable inputs for the cutoff parameter for the form factor, we find that the interaction of the OSE and OOE models provides attraction which may form the heavy molecular states, like the S-wave $\Lambda_cD$, $\Lambda_c\Lambda_c$, and $\Lambda_c\bar{\Lambda}_c$ states.

As a by-product, we also discuss the mass dependence of the binding energy. We have explicitly shown that heavier systems are more likely to accommodate various molecular states such as S-wave $\Lambda_bD(\bar{D})$, $\Lambda_cB(\bar{B})$, $\Lambda_bB(\bar{B})$, $\Lambda_c\Lambda(\bar{\Lambda})$, $\Lambda_b\Lambda(\bar{\Lambda})$, $\Lambda_b\Lambda_c(\bar{\Lambda}_c)$, and $\Lambda_b\Lambda_b(\bar{\Lambda}_b)$ states. Finally, the interaction between a heavy baryon $\Lambda_{c,b}$ and one nucleon has been investigated. In our calculation, there can exist S-wave $\Lambda_cN(\bar{N})$ and $\Lambda_bN(\bar{N})$ molecular states.

%\appendix

\vfill
%\newpage

\section*{ACKNOWLEDGMENTS}

This project is partly supported by the National Natural Science Foundation of China under Grants No. 11222547, No. 11175073, and No. 11647301, and the Fundamental Research Funds for the Central Universities. X. L. is also supported in part by the National Program for Support of Top-notch Young Professionals. A. H. is supported in part by Grants-in-Aid for Scientific Research [Grant No. JPK05441(c)].


\begin{thebibliography}{99}

%\cite{Choi:2003ue}
\bibitem{Choi:2003ue}
  S.~K.~Choi {\it et al.} [Belle Collaboration],
  Observation of a narrow charmonium - like state in exclusive $B^{\pm} \to K^{\pm} \pi^+ \pi^- J / \psi$ decays,
  Phys.\ Rev.\ Lett.\  {\bf 91}, 262001 (2003).
 % [hep-ex/0309032].
  %%CITATION = doi:10.1103/PhysRevLett.91.262001;%%
  %1346 citations counted in INSPIRE as of 27 Jun 2017

%\cite{Nakano:2003qx}
\bibitem{Nakano:2003qx}
  T.~Nakano {\it et al.} [LEPS Collaboration],
  Evidence for a narrow $S = +1$ baryon resonance in photoproduction from the neutron,
  Phys.\ Rev.\ Lett.\  {\bf 91}, 012002 (2003).
%  [hep-ex/0301020].
  %%CITATION = doi:10.1103/PhysRevLett.91.012002;%%
  %1064 citations counted in INSPIRE as of 27 Jun 2017

%\cite{Battaglieri:2005er}
\bibitem{Battaglieri:2005er}
  M.~Battaglieri {\it et al.} [CLAS Collaboration],
  Search for $\Theta+(1540)$ pentaquark in high statistics measurement of $\gamma p \to \bar{K}^0 K^+ n$ at CLAS,
  Phys.\ Rev.\ Lett.\  {\bf 96}, 042001 (2006)
%  doi:10.1103/PhysRevLett.96.042001
%  [hep-ex/0510061].
  %%CITATION = doi:10.1103/PhysRevLett.96.042001;%%
  %106 citations counted in INSPIRE as of 23 Sep 2017

%\cite{McKinnon:2006zv}
\bibitem{McKinnon:2006zv}
  B.~McKinnon {\it et al.} [CLAS Collaboration],
  Search for the Theta+ pentaquark in the reaction $\gamma d \to p K^- K^+ n$,
  Phys.\ Rev.\ Lett.\  {\bf 96}, 212001 (2006)
 % doi:10.1103/PhysRevLett.96.212001
 % [hep-ex/0603028].
  %%CITATION = doi:10.1103/PhysRevLett.96.212001;%%
  %116 citations counted in INSPIRE as of 23 Sep 2017

%\cite{Belle:2011aa}
\bibitem{Belle:2011aa}
  A.~Bondar {\it et al.} [Belle Collaboration],
  Observation of two charged bottomonium-like resonances in $\Upsilon(5S)$ decays,
  Phys.\ Rev.\ Lett.\  {\bf 108}, 122001 (2012).
%  [arXiv:1110.2251 [hep-ex]].
  %%CITATION = doi:10.1103/PhysRevLett.108.122001;%%
  %361 citations counted in INSPIRE as of 27 Jun 2017

%\cite{Aaij:2015tga}
\bibitem{Aaij:2015tga}
  R.~Aaij {\it et al.} [LHCb Collaboration],
  Observation of $J/\psi$ Resonances Consistent with Pentaquark States in $\Lambda_b^0\rightarrow J/\psi K^-p$ Decays,
  Phys.\ Rev.\ Lett.\  {\bf 115}, 072001 (2015).
  %doi:10.1103/PhysRevLett.115.072001
%  [arXiv:1507.03414 [hep-ex]].
  %%CITATION = doi:10.1103/PhysRevLett.115.072001;%%
  %83 citations counted in INSPIRE as of 24 Nov 2015

%\cite{Chen:2016qju}
\bibitem{Chen:2016qju}
  H.~X.~Chen, W.~Chen, X.~Liu and S.~L.~Zhu,
  The hidden-charm pentaquark and tetraquark states,
  Phys.\ Rept.\  {\bf 639}, 1 (2016).
 % doi:10.1016/j.physrep.2016.05.004
 % [arXiv:1601.02092 [hep-ph]].
  %%CITATION = doi:10.1016/j.physrep.2016.05.004;%%
  %175 citations counted in INSPIRE as of 19 Jul 2017

%\cite{Liu:2013waa}
\bibitem{Liu:2013waa}
  X.~Liu,
  An overview of $XYZ$ new particles,
  Chin.\ Sci.\ Bull.\  {\bf 59}, 3815 (2014).
  %doi:10.1007/s11434-014-0407-2
 % [arXiv:1312.7408 [hep-ph]].
  %%CITATION = doi:10.1007/s11434-014-0407-2;%%
  %67 citations counted in INSPIRE as of 19 Jul 2017

%\cite{Hosaka:2016pey}
\bibitem{Hosaka:2016pey}
  A.~Hosaka, T.~Iijima, K.~Miyabayashi, Y.~Sakai and S.~Yasui,
  Exotic hadrons with heavy flavors: X, Y, Z, and related states,
  PTEP {\bf 2016}, 062C01 (2016)
 % doi:10.1093/ptep/ptw045
 % [arXiv:1603.09229 [hep-ph]].
  %%CITATION = doi:10.1093/ptep/ptw045;%%
  %29 citations counted in INSPIRE as of 25 Jul 2017

%\cite{Liu:2008fh}
\bibitem{Liu:2008fh}
  Y.~R.~Liu, X.~Liu, W.~Z.~Deng and S.~L.~Zhu,
  Is $X(3872) $ Really a Molecular State?,
  Eur.\ Phys.\ J.\ C {\bf 56}, 63 (2008).
  %doi:10.1140/epjc/s10052-008-0640-4
  %[arXiv:0801.3540 [hep-ph]].
  %%CITATION = doi:10.1140/epjc/s10052-008-0640-4;%%
  %106 citations counted in INSPIRE as of 21 Jan 2017

%\cite{Thomas:2008ja}
\bibitem{Thomas:2008ja}
  C.~E.~Thomas and F.~E.~Close,
  Is $X(3872)$ a molecule?,
  Phys.\ Rev.\ D {\bf 78}, 034007 (2008).
  %doi:10.1103/PhysRevD.78.034007
  %[arXiv:0805.3653 [hep-ph]].
  %%CITATION = doi:10.1103/PhysRevD.78.034007;%%
  %100 citations counted in INSPIRE as of 21 Jan 2017

%\cite{Lee:2009hy}
\bibitem{Lee:2009hy}
  I.~W.~Lee, A.~Faessler, T.~Gutsche and V.~E.~Lyubovitskij,
  $X(3872)$ as a molecular $D\bar{D}^*$ state in a potential model,
  Phys.\ Rev.\ D {\bf 80}, 094005 (2009).
 % doi:10.1103/PhysRevD.80.094005
  %[arXiv:0910.1009 [hep-ph]].
  %%CITATION = doi:10.1103/PhysRevD.80.094005;%%
  %61 citations counted in INSPIRE as of 21 Jan 2017

%\cite{Li:2012cs}
\bibitem{Li:2012cs}
  N.~Li and S.~L.~Zhu,
  Isospin breaking, Coupled-channel effects and Diagnosis of $X(3872)$,
  Phys.\ Rev.\ D {\bf 86}, 074022 (2012).
  %doi:10.1103/PhysRevD.86.074022
  %[arXiv:1207.3954 [hep-ph]].
  %%CITATION = doi:10.1103/PhysRevD.86.074022;%%
  %21 citations counted in INSPIRE as of 21 Jan 2017

%\cite{Sun:2012zzd}
\bibitem{Sun:2012zzd}
  Z.~F.~Sun, Z.~G.~Luo, J.~He, X.~Liu and S.~L.~Zhu,
  A note on the $B^*\bar{B}$, $B^*\bar{B}^*$, $D^*\bar{D}$, and $D^*\bar{D}^*$ molecular states,
  Chin.\ Phys.\ C {\bf 36}, 194 (2012).
 % doi:10.1088/1674-1137/36/3/002
  %%CITATION = doi:10.1088/1674-1137/36/3/002;%%
  %26 citations counted in INSPIRE as of 21 Jan 2017

%\cite{Zhao:2014gqa}
\bibitem{Zhao:2014gqa}
  L.~Zhao, L.~Ma and S.~L.~Zhu,
  Spin-orbit force, recoil corrections, and possible $B \bar{B}^{*}$ and $D \bar{D}^{*}$  molecular states,
  Phys.\ Rev.\ D {\bf 89}, no. 9, 094026 (2014).
  %doi:10.1103/PhysRevD.89.094026
  %[arXiv:1403.4043 [hep-ph]].
  %%CITATION = doi:10.1103/PhysRevD.89.094026;%%
  %11 citations counted in INSPIRE as of 21 Jan 2017

%\cite{Ikeda:2016zwx}
\bibitem{Ikeda:2016zwx}
  Y.~Ikeda {\it et al.} [HAL QCD Collaboration],
  Fate of the Tetraquark Candidate $Z_c$(3900) from Lattice QCD,
  Phys.\ Rev.\ Lett.\  {\bf 117}, 242001 (2016)
 % doi:10.1103/PhysRevLett.117.242001
 % [arXiv:1602.03465 [hep-lat]].
  %%CITATION = doi:10.1103/PhysRevLett.117.242001;%%
  %21 citations counted in INSPIRE as of 25 Jul 2017

%\cite{Tornqvist:1993vu}
\bibitem{Tornqvist:1993vu}
  N.~A.~Tornqvist,
  On deusons or deuteron - like meson meson bound states,
  Nuovo Cim.\ A {\bf 107}, 2471 (1994).
%  doi:10.1007/BF02734018
%  [hep-ph/9310225].
  %%CITATION = doi:10.1007/BF02734018;%%
  %33 citations counted in INSPIRE as of 01 Feb 2017

%\cite{Tornqvist:1993ng}
\bibitem{Tornqvist:1993ng}
  N.~A.~Tornqvist,
  From the deuteron to deusons, an analysis of deuteron-like meson meson bound states,
  Z.\ Phys.\ C {\bf 61}, 525 (1994).
  %doi:10.1007/BF01413192
  %[hep-ph/9310247].
  %%CITATION = doi:10.1007/BF01413192;%%
  %298 citations counted in INSPIRE as of 01 Feb 2017

%\cite{Wu:2010jy}
\bibitem{Wu:2010jy}
  J.~J.~Wu, R.~Molina, E.~Oset and B.~S.~Zou,
  Prediction of narrow $N^*$ and $\Lambda^*$ resonances with hidden charm above 4 GeV,
  Phys.\ Rev.\ Lett.\  {\bf 105}, 232001 (2010).
 % doi:10.1103/PhysRevLett.105.232001
 % [arXiv:1007.0573 [nucl-th]].
  %%CITATION = doi:10.1103/PhysRevLett.105.232001;%%
  %131 citations counted in INSPIRE as of 14 Feb 2017

%\cite{Karliner:2016ith}
\bibitem{Karliner:2016ith}
  M.~Karliner and J.~L.~Rosner,
  Exotic resonances due to $\eta$ exchange,
  Nucl.\ Phys.\ A {\bf 954}, 365 (2016).
 % doi:10.1016/j.nuclphysa.2016.03.057
 % [arXiv:1601.00565 [hep-ph]].
  %%CITATION = doi:10.1016/j.nuclphysa.2016.03.057;%%
  %12 citations counted in INSPIRE as of 14 Feb 2017

%\cite{Chen:2016ryt}
\bibitem{Chen:2016ryt}
  R.~Chen, J.~He and X.~Liu,
  Possible strange hidden-charm pentaquarks from $\Sigma_c^{(*)}\bar{D}_s^*$ and $\Xi^{(',*)}_c\bar{D}^*$ interactions,
  arXiv:1609.03235 [hep-ph].
  %%CITATION = ARXIV:1609.03235;%%
  %2 citations counted in INSPIRE as of 14 Feb 2017



%\cite{Meguro:2011nr}
\bibitem{Meguro:2011nr}
  W.~Meguro, Y.~R.~Liu and M.~Oka,
  Possible $\Lambda_c\Lambda_c$ molecular bound state,
  Phys.\ Lett.\ B {\bf 704}, 547 (2011)
  %doi:10.1016/j.physletb.2011.09.088
  %[arXiv:1105.3693 [hep-ph]].
  %%CITATION = doi:10.1016/j.physletb.2011.09.088;%%
  %27 citations counted in INSPIRE as of 27 Jul 2017

%\cite{Li:2012bt}
\bibitem{Li:2012bt}
  N.~Li and S.~L.~Zhu,
  Hadronic Molecular States Composed of Heavy Flavor Baryons,
  Phys.\ Rev.\ D {\bf 86}, 014020 (2012)
  %doi:10.1103/PhysRevD.86.014020
  %[arXiv:1204.3364 [hep-ph]].
  %%CITATION = doi:10.1103/PhysRevD.86.014020;%%
  %14 citations counted in INSPIRE as of 27 Jul 2017

%\cite{Lee:2011rka}
\bibitem{Lee:2011rka}
  N.~Lee, Z.~G.~Luo, X.~L.~Chen and S.~L.~Zhu,
  Possible Deuteron-like Molecular States Composed of Heavy Baryons,
  Phys.\ Rev.\ D {\bf 84}, 014031 (2011)
 % doi:10.1103/PhysRevD.84.014031
  %[arXiv:1104.4257 [hep-ph]].
  %%CITATION = doi:10.1103/PhysRevD.84.014031;%%
  %21 citations counted in INSPIRE as of 27 Jul 2017


%\cite{Riska:1999fn}
\bibitem{Riska:1999fn}
  D.~O.~Riska and G.~E.~Brown,
  Two pion exchange interaction between constituent quarks,
  Nucl.\ Phys.\ A {\bf 653}, 251 (1999).
 % doi:10.1016/S0375-9474(99)00260-2
 % [hep-ph/9902319].
  %%CITATION = doi:10.1016/S0375-9474(99)00260-2;%%
  %40 citations counted in INSPIRE as of 16 Feb 2017

%\cite{Rijken:1998yy}
\bibitem{Rijken:1998yy}
  T.~A.~Rijken, V.~G.~J.~Stoks and Y.~Yamamoto,
  Soft core hyperon - nucleon potentials,
  Phys.\ Rev.\ C {\bf 59}, 21 (1999).
  %doi:10.1103/PhysRevC.59.21
  %[nucl-th/9807082].
  %%CITATION = doi:10.1103/PhysRevC.59.21;%%
  %433 citations counted in INSPIRE as of 20 Feb 2017

%\cite{Riska:2000gd}
\bibitem{Riska:2000gd}
  D.~O.~Riska and G.~E.~Brown,
  Nucleon resonance transition couplings to vector mesons,
  Nucl.\ Phys.\ A {\bf 679}, 577 (2001).
  %doi:10.1016/S0375-9474(00)00362-6
  %[nucl-th/0005049].
  %%CITATION = doi:10.1016/S0375-9474(00)00362-6;%%
  %89 citations counted in INSPIRE as of 16 Feb 2017

%\cite{Klempt:2002ap}
\bibitem{Klempt:2002ap}
  E.~Klempt, F.~Bradamante, A.~Martin and J.~M.~Richard,
  Antinucleon nucleon interaction at low energy: Scattering and protonium,
  Phys.\ Rept.\  {\bf 368}, 119 (2002).
 % doi:10.1016/S0370-1573(02)00144-8
  %%CITATION = doi:10.1016/S0370-1573(02)00144-8;%%
  %107 citations counted in INSPIRE as of 20 Feb 2017


%\cite{Olive:2016xmw}
\bibitem{Olive:2016xmw}
  C.~Patrignani {\it et al.} [Particle Data Group],
  Review of Particle Physics,
  Chin.\ Phys.\ C {\bf 40}, 10, 100001 (2016).
 % doi:10.1088/1674-1137/40/10/100001
  %%CITATION = doi:10.1088/1674-1137/40/10/100001;%%
  %728 citations counted in INSPIRE as of 27 Apr 2017


%\cite{Dover:1977jw}
\bibitem{Dover:1977jw}
  C.~B.~Dover and S.~H.~Kahana,
  Possibility of Charmed Hypernuclei,
  Phys.\ Rev.\ Lett.\  {\bf 39}, 1506 (1977).
  %doi:10.1103/PhysRevLett.39.1506
  %%CITATION = doi:10.1103/PhysRevLett.39.1506;%%
  %67 citations counted in INSPIRE as of 27 Jul 2017

%\cite{Liu:2011xc}
\bibitem{Liu:2011xc}
  Y.~R.~Liu and M.~Oka,
  $\Lambda_c N$ bound states revisited,
  Phys.\ Rev.\ D {\bf 85}, 014015 (2012)
 % doi:10.1103/PhysRevD.85.014015
 % [arXiv:1103.4624 [hep-ph]].
  %%CITATION = doi:10.1103/PhysRevD.85.014015;%%
  %43 citations counted in INSPIRE as of 27 Jul 2017

 %\cite{Meng:2017udf}
\bibitem{Meng:2017udf}
  L.~Meng, N.~Li and S.~l.~Zhu,
  Possible hadronic molecules composed of the doubly charmed baryon and nucleon,
  arXiv:1707.03598 [hep-ph].
  %%CITATION = ARXIV:1707.03598;%%
  %1 citations counted in INSPIRE as of 27 Jul 2017


%\cite{Maeda:2015hxa}
\bibitem{Maeda:2015hxa}
  S.~Maeda, M.~Oka, A.~Yokota, E.~Hiyama and Y.~R.~Liu,
  A model of charmed baryon¨Cnucleon potential and two- and three-body bound states with charmed baryon,
  PTEP {\bf 2016}, 023D02 (2016).
 % doi:10.1093/ptep/ptv194
  %[arXiv:1509.02445 [nucl-th]].
  %%CITATION = doi:10.1093/ptep/ptv194;%%
  %13 citations counted in INSPIRE as of 28 Apr 2017


%\cite{Machleidt:1987hj}
\bibitem{Machleidt:1987hj}
  R.~Machleidt, K.~Holinde and C.~Elster,
  The Bonn Meson Exchange Model for the Nucleon Nucleon Interaction,
  Phys.\ Rept.\  {\bf 149}, 1 (1987).
 % doi:10.1016/S0370-1573(87)80002-9
  %%CITATION = doi:10.1016/S0370-1573(87)80002-9;%%
  %2127 citations counted in INSPIRE as of 25 Sep 2017

%\cite{Backman:1984sx}
\bibitem{Backman:1984sx}
  S.~O.~Backman, G.~E.~Brown and J.~A.~Niskanen,
  The Nucleon Nucleon Interaction And The Nuclear Many Body Problem,
  Phys.\ Rept.\  {\bf 124}, 1 (1985).
 % doi:10.1016/0370-1573(85)90108-5
  %%CITATION = doi:10.1016/0370-1573(85)90108-5;%%
  %153 citations counted in INSPIRE as of 25 Sep 2017




\end{thebibliography}
\end{document}